# Concrete and Lead Shielding Requirements for PET Facilities

V. Steiner[1], A. Malki[1], T. Ben-Yehuda[1], M. Moinester[2*]

[1]Department of Nuclear Medicine, Sheba Medical Center, 5262000 Ramat Gan, Israel
[2]School of Physics and Astronomy, Tel Aviv University, 69978 Tel Aviv, Israel

*Corresponding author's email: murraym@tauex.tau.ac.il

**Article Type:** Research article

**Complete Details of Each Author**

First Author's Full Name: Victor Steiner
Highest Qualification: Ph.D.
Department: Department of Nuclear Medicine
Post/Rank: Consultant
Affiliation: Sheba Medical Center, 5262000 Ramat Gan, Israel
email id: vikilica@gmail.com
ORCID: https://orcid.org/0009-0006-5697-7277

Second Author's Full Name: Avi Malki
Highest Qualification: Ph.D.
Department: Department of Nuclear Medicine
Post/Rank: Consultant
Affiliation: Sheba Medical Center, 5262000 Ramat Gan, Israel
email id: Avi.Malki@sheba.health.gov.il
ORCID: NA

Third Author's Full Name: Tzafrir Ben-Yehuda
Highest Qualification: MSc. Biomedical Engineer
Department: Department of Nuclear Medicine
Post/Rank: Consultant
Affiliation: Sheba Medical Center, 5262000 Ramat Gan, Israel
email id: Tzafrir.Benyehuda@sheba.health.gov.il
ORCID: NA

Fourth Author's Full Name: Murray Moinester
Highest Qualification: Ph.D.
Department: School of Physics and Astronomy
Post/Rank: Emeritus Professor
Affiliation: Tel Aviv University, 69978 Tel Aviv, Israel
email id: murraym@tauex.tau.ac.il
web page: murraymoinester.com
ORCID: https://orcid.org/0000-0001-8764-5618

## Abstract

We investigate shielding thickness requirements in concrete and lead walls for positron emission tomography (PET) facilities. $F^{18}$, the most commonly used PET radiotracer, emits two back-to-back 511 keV photons, necessitating effective shielding to protect hospital staff while patients are in treatment rooms and during their hospital stay, whether stationary or walking. Photon transmission measurements were conducted through a standard Israeli B30 concrete wall (3 m high, 20 cm thick, density 2.30 g/cm$^3$) using photons from an $F^{18}$ source with a 24° FWHM dose aperture opening angle. A narrow-beam transmission coefficient of T = (3.0 ± 1.0)% was recorded with the source positioned 0.05 m from the wall, and the detector at distances of 0.05-3 m on the opposite side. When the source is 3 meters from the wall, photons within a 0.64 m radius circular disk (wide-beam) strikes the wall. This produces a dose "buildup" effect, whereby photons striking within the disk reach the dosimeter after Compton scattering within the wall. The wide-beam transmission coefficient was measured as T = (8.8 ± 1.8)%, corresponding to a buildup factor B = 4.0 ± 0.8, consistent with Monte Carlo (MC) simulations B = 3.9 ± 0.6, validating the simulation reliability for this specific setup. However, we caution against assuming universal validation for all MC algorithms. Furthermore, reliance solely on thick concrete walls presents construction challenges due to weight, space and cost inefficiency. Since ordinary concrete effectively interacts only through Compton scattering, while lead walls interact via the photoelectric effect, lead is preferable for shielding. A 3.3 cm lead wall results in <0.5% transmission, and lead sheets can be supported by a 5 cm gypsum wall, optimizing shielding design while addressing space and cost considerations.

**Keywords:** PET radiation shielding, buildup factors, Monte Carlo simulations

## 1 Introduction

This study focuses on protective concrete shielding walls in positron emission tomography (PET) facilities, drawing parallels with those at Sheba Hospital in Israel [1]. The most prevalent radiotracer for PET scans is the radioisotope $F^{18}$. This isotope has a physical half-life of 1.8 hours and an effective lifetime of 1.4 hours, which encompasses both physical decay and biological clearance. Its primary 97% decay mode involves the emission of a positron with a mean energy of 250 keV. Upon encountering tissue, the positron travels approximately 1.2 mm before annihilating with an electron, producing two 511 keV gamma rays (photons) emitted isotopically in opposite directions. Following the administration of a typical dose of 10 mCi, the patient's $F^{18}$ radioactivity decreases to about 14% after four hours.

To ensure the safety of hospital staff from radiation exposure during a patient's stay, shielding walls are essential in PET facilities [2]. We conducted photon transmission measurements through constructed concrete shielding walls to establish the necessary concrete thickness to comply with radiation protection regulations. The relationship between photon transmission and various parameters can be expressed via the transmission coefficient T, defined as [3]:

$$T = I/I_0 = B(E, Z, x)\, exp(-\mu x). \qquad (1)$$

Here, I and $I_0$ represent the transmitted and incident photon intensities, respectively. The dose buildup factor B(E, Z, x) quantifies how much additional dose is delivered beyond what would be expected from simple exponential attenuation alone. The reason is that scattered photons contribute to the dose received by an object or tissue or dosemeter beyond those which pass directly through unscattered. The buildup factor B depends on photon energy E, average atomic number Z, shielding thickness x, and beam geometry. The linear attenuation coefficient μ enters through the exponential attenuation term. Narrow and wide photon beams interact differently with shielding materials. A narrow beam allows only photons traveling along the beam axis to reach a dosemeter that is centered on the beam axis, while a wide beam also enables off-axis photons to scatter from the shielding wall towards the detector.

For typical PET energies around 511 keV, attenuation occurs through combinations of photoelectric absorption and Compton scattering. The buildup factor is generally estimated through Monte Carlo (MC) simulation calculations, and tabulated according to Z, E, x. The thickness x in these tables is often measured in units of $\lambda = 1/\mu$ mean free paths (mfp). The mass attenuation coefficient value $\mu_m = 0.0833$ $cm^2/g$ ($\mu$=0.196 $cm^{-1}$, $\lambda$=5.10 cm), reported by Stankovic et al. [4] for concrete is taken for all calculations. It is close to the NIST value [5].

To determine the buildup factor for concrete at photon energies around 511 keV, we conducted photon transmission measurements through a wall constructed with standard Israeli B30 concrete (3 meters high, 20 cm thick, density 2.30 g/cm³, 30 MPa compressive strength, Z ~10-12, A ~ 20-25) [6]. These measurements employed photons from an $F^{18}$ source. Both narrow-beam and wide-beam transmission coefficients and associated buildup factors were measured with the source placed 0.05 m and 3.0 m from the wall, respectively; while the detector was positioned at distances ranging from 0.05 to 3 m on the opposite side. All measurements were done at the Sheba Hospital PET facility construction site. There was no ceiling, no nearby walls, and no other objects present besides the floor; thereby minimizing scattering backgrounds. A photon beam was directed at a constructed shielding wall. The beam exited an opening in the housing of the source, which was otherwise completely shielded. Following measurements of the transmission coefficient through the concrete wall, we compare the associated buildup factors with available Monte Carlo simulation calculations.

Note however that relying on thick concrete walls poses challenges related to construction weight, cost and space inefficiency. Standard concrete interacts weakly with photons mainly via Compton scattering. Lead shielding walls are shown to be a superior alternative, considering that they interact with photons much more strongly via the photoelectric effect.

## 2 Research Methodology

### 2.1 $F^{18}$ Source

The measurements employed photons emitted from a liquid Fludeoxyglucose (FDG) $F^{18}$ source [7]. A schematic drawing of the source assembly is shown in Fig. 1. The source activity was 23.4 mCi (865 MBq) at the start of measurements. Since $F^{18}$ decays mainly by emitting two 511 keV photons back-to-back, one therefore expects $8.65\times10^8$ emitted photons per second into the forward hemisphere. The FDG was held in a 20 ml glass vial in a saline (NaCl) solution having density ~ 1 g/cm³. The vial was secured inside a 2 cm thick tungsten container, which itself was secured inside a 1 cm thick lead-plated transportation box. The thick tungsten walls allowed photons to exit only through a geometric opening angle of 32°, through a 0.3 cm thick cylindrical tungsten collimator, having diameter 2 cm and a central 1 cm diameter aperture. Aside from a small component of back scattering, forward emitted photons exit the source either through the collimator's aperture or by transmission through its 0.3 cm thickness. The central aperture is expected to transmit mainly the non-interacting 511 keV photons emitted from about 20% of the source volume. The rest are absorbed in the tungsten container. The overall transmission coefficient out of the transportation box was 0.1%, ensuring a low radiation field of about 1 μSv/h at 1 m distance from the box. Only about 80% of the vial was filled. This feature lowered the emission probability of the source when rotated 90° for measurement in the horizontal direction. The reason is that the volume of an imaginary 2 cm diameter cylinder centered inside the vial, although aligned with the 2 cm diameter of the collimator, is not completely filled with liquid. Only photons emitted from source liquid within this cylinder are geometrically positioned to efficiently exit through the collimator.

At 511 keV, the photon interaction cross sections with tungsten, for coherent (Rayleigh), incoherent (Compton) and photoelectric effects are 2.9, 21.0 and 18.2 barn, respectively [5]. The relative interaction probabilities are 7%, 50%, 43%. In coherent scattering, the angular distribution is isotropic and the energy does not change with scattering angle. In Compton interactions, scattered photons have energies between 170 and 511 keV; the larger the scattering angle, the lower the energy. The angular distribution is forward peaked, with 75% of the photons emitted in the forward hemisphere, based on the Compton (Klein-Nishima) photon angular distribution at 511 keV. In photoelectric events, the photons are absorbed.

The mean free path (mfp) of photon absorption in tungsten at 511 keV is 0.38 cm [5]. As a result, the 0.3 cm thick tungsten collimator sheet transmits about 45% ($e^{-0.3/0.42}$) of the incident direct narrow-beam (511 keV) photons. Some photons will exit with reduced energy because of Compton scattering. The mfp for absorption in tungsten decreases sharply with decreasing photon energy. Photons emitted from the source are therefore expected to have energies primarily in the range 300 to 511 keV, with a strong peak at 511 keV.

Based on source geometry, roughly 30% of the $8.65\times10^8$ photons per second forward emitted by $F^{18}$ exit through the collimator. The beam intensity is expected to decrease from small to large angles; and should be comprised of ~511 keV photons at small angles, and lower energy photons at larger angles. We confirmed this expected beam structure with a NaI (Tl) detector.

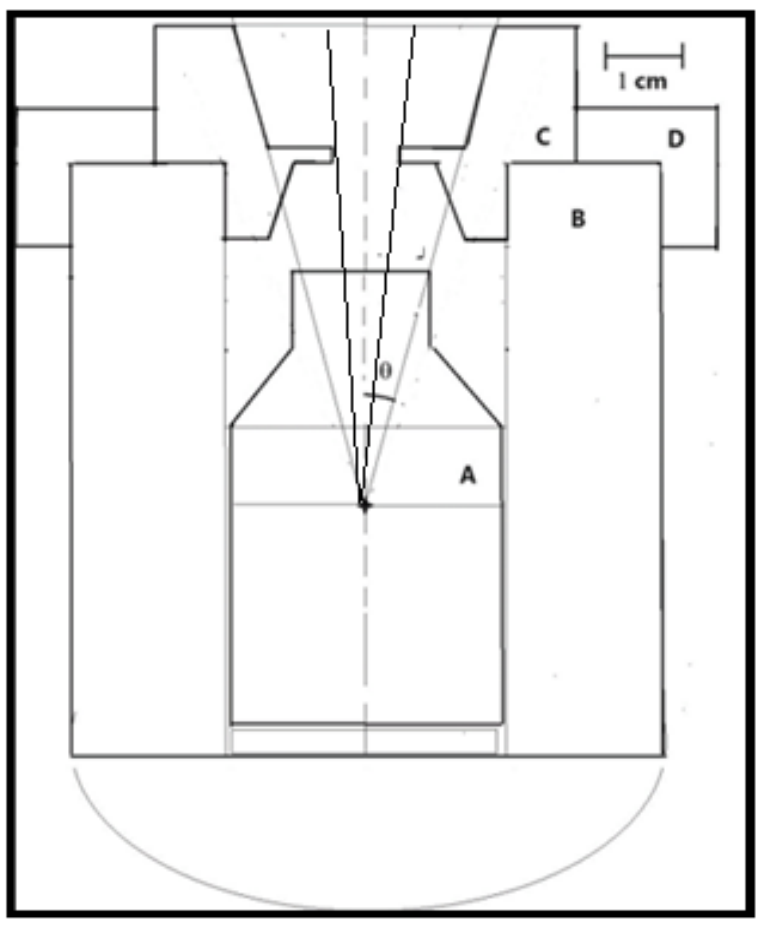

Fig.1: Scale drawing (see upper right corner) of the $F^{18}$ source, showing A: Glass vial containing liquid FDG, B: Tungsten container, C: Tungsten collimator, D: Aluminum support positioning the collimator onto the tungsten container. The geometric beam angle $\theta = 16°$ (half opening angle) is shown with respect to the vial's central axis.

## 2.2 Beam profile

The geometry of the beam was checked by direct measurement, as shown in Fig. 2a. The source was positioned far from walls, and dose measurements were carried out 50 cm from the source, as shown in the figure. Since the detector measures dose rate, it is sensitive to the energy of detected photons, and is therefore sensitive to the angular distribution of scattered photons. Dose measurements in μSv/h were carried out with the Atomtex dosemeter [8] at a fixed distance y = 50 cm, over the horizontal distance range x = ± 50 cm, as shown in the figure. Fig. 2b shows a Gaussian fit to the data. The normal distribution chi-square fit gives $\theta_{rms} = 10.3°$. The effective beam angle $\theta_s$ is then given by:

$$\theta_s = \frac{\mathrm{FWHM}}{2} = \frac{2.35 \cdot \theta(\mathrm{rms})}{2} = 1.17 \cdot \theta(\mathrm{rms}) = 12^0 \qquad \textbf{(2)}$$

That is, the photons are effectively emitted from the source into a cone having a 24° FWHM dose aperture angle, smaller than the geometric cone angle of 32°. The beam contains photons emitted directly through the collimator aperture at an angle of 0-5 degrees with maximum energy 511 keV; as well as photons Compton scattered in the collimator or in the source container with energies in the range 170-511 keV. The photon beam energy is maximal on the beam axis and decreases with increasing beam angle. Since radiation dose is proportional to the amount of energy delivered by the photons, the lower energy photons deliver less dose. Note that for measuring a transmission coefficient, the beam does not have to be only 511 keV, but rather similar to the photon energy spectrum emitted by a PET facility patient, stationary or moving.

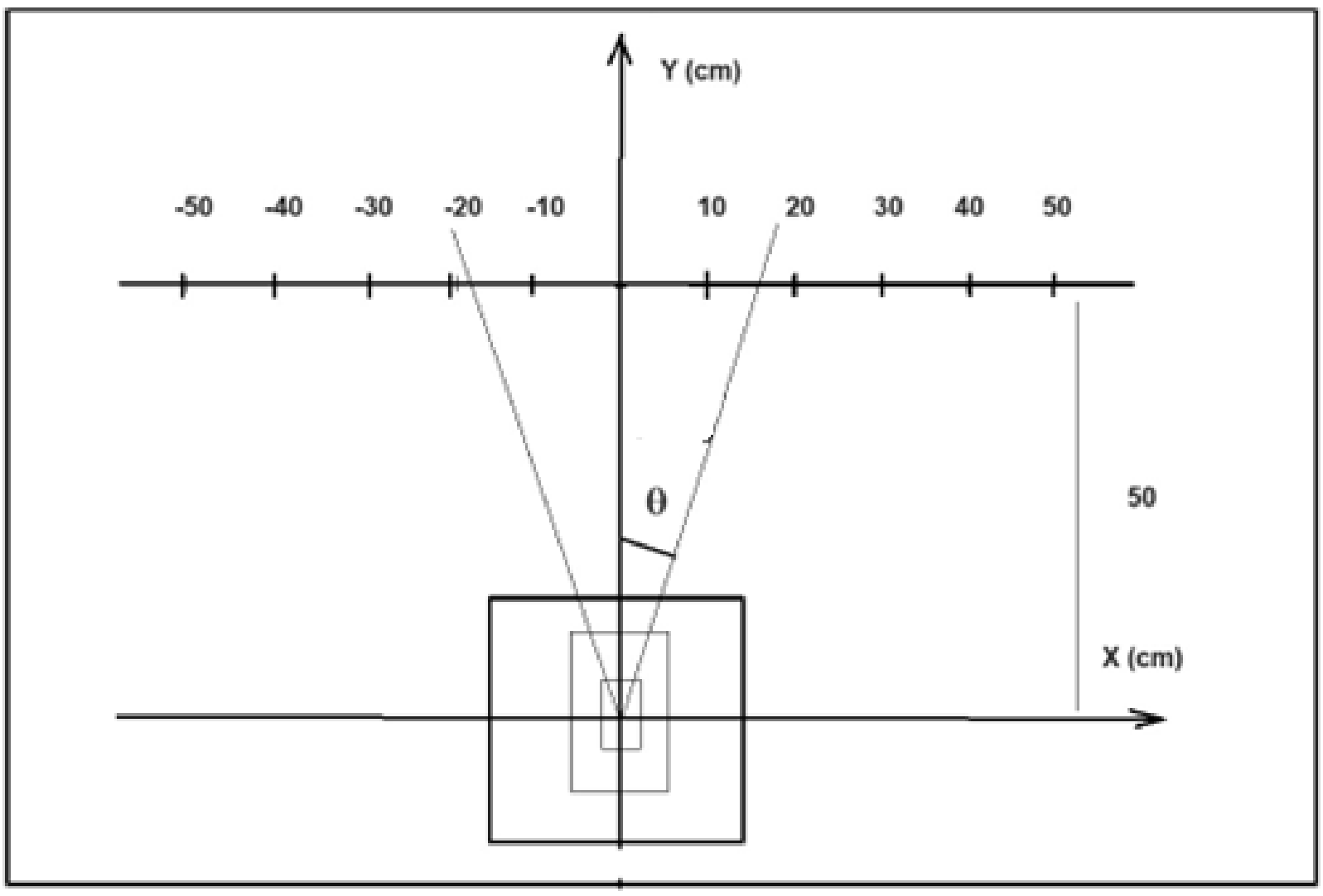

Fig. 2a: Setup for measuring the beam profile

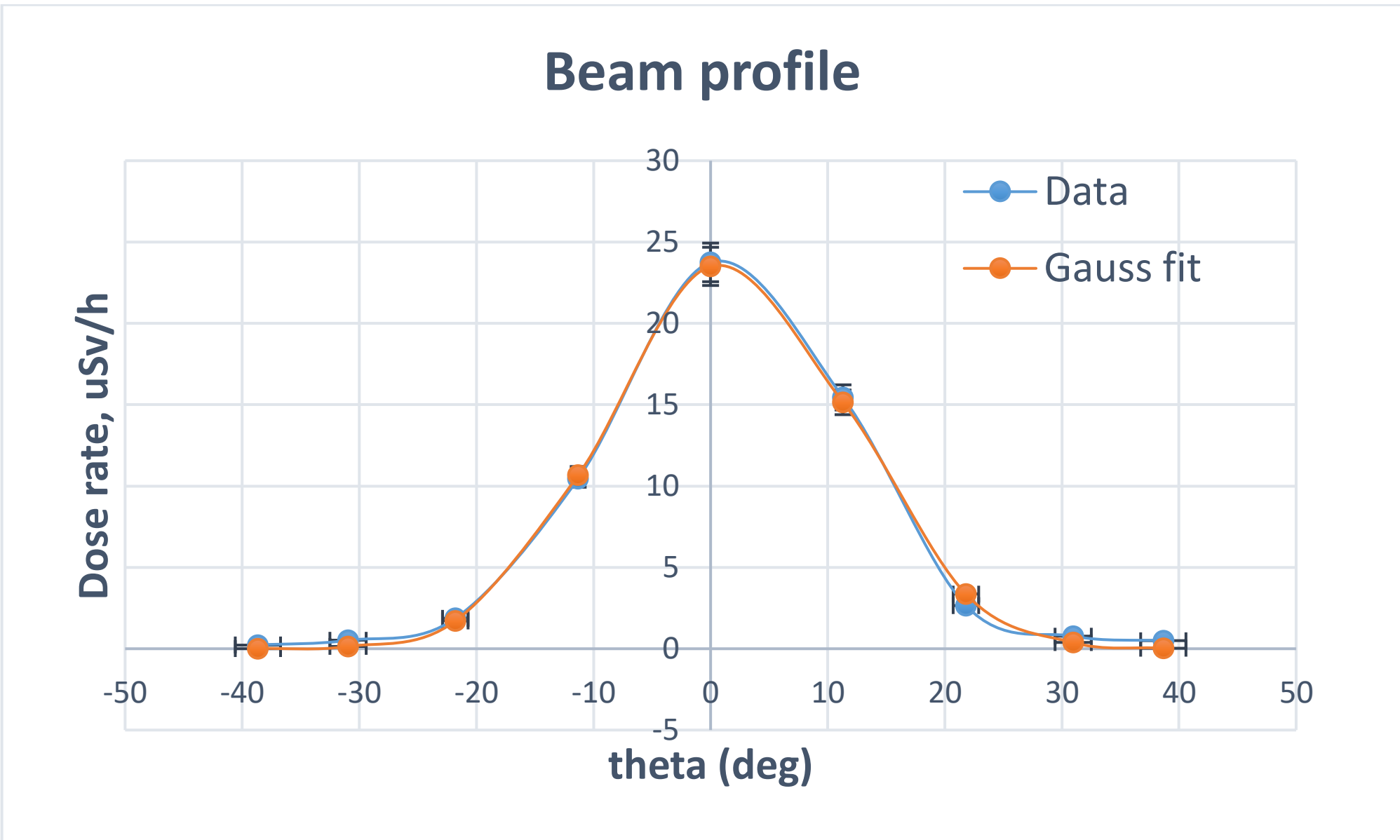

Fig. 2b: Radiation dose rate measured by the Atomex dose meter as a function of the angle θ relative to the beam axis, together with a chi-square fit to a Gaussian function.

## 2.3 Atomtex AT1123 Radiation Survey Meter (Dosimeter)

This Atomtex dosimeter [8] uses a plastic scintillator (3 cm diameter, 1.5 cm thick) to detect a high energy photon (511 keV) by converting it into numerous low energy photons (visible light) in the range 1.6 – 3.3 eV through the process of scintillation. The light flash intensity is proportional to the energy deposited by the incident photon. The visible light photons release electrons from a photomultiplier tube photocathode. The photomultiplier then converts this low intensity pulse of electrons into a high intensity pulse of electric charge, whose integrated charge is proportional to the incident photon energy. This charge provides a measure of the photon energy. Measuring their magnitude determines the rate of energy deposition by the beam photons [9]. The AT1123 dosemeter was calibrated and certified annually by the Israeli Radiation Protection Authority by comparison to a calibrated NIST high pressure ionization chamber, with calibration accuracy of about 10% [10].

## 2.4 Measurement Method

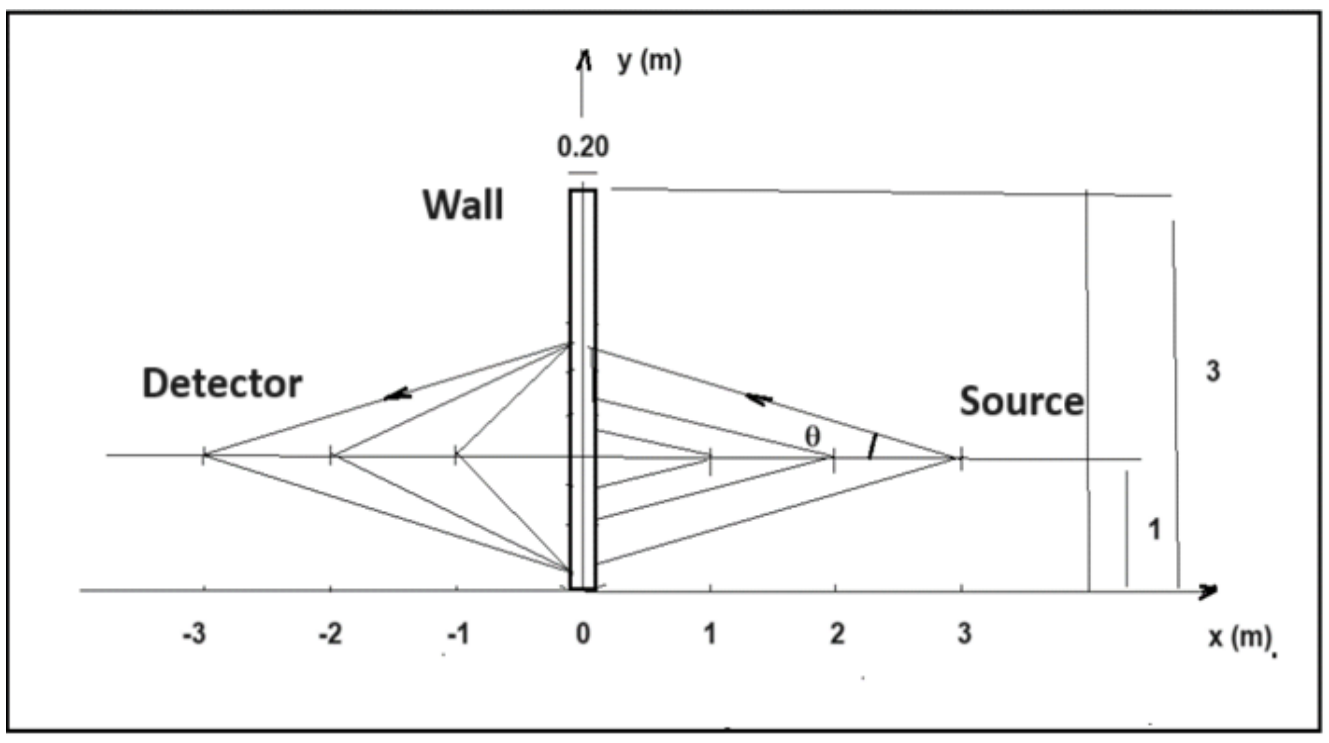

Fig. 3: Experimental Setup showing variable distance options.

All measurements were carried out with the same source. Fig. 3 shows the experimental setup. The source, positioned 3 meters from the wall, yielded a 0.64 m radius circular beam disk (wide-beam) on the concrete wall. The measurements were carried out at the Sheba Hospital PET facility construction site on an isolated wall. without ceiling. This setup avoided back-scattering effects. Dose rates were measured by an Atomtex AT1123 Radiation Survey Meter [8], positioned between 0.05 to 3 meters from the far side of the wall, and these measurements were averaged. The measurement uncertainty was taken as the standard deviation of the measurements. Distances were measured by a tape measure.

The measurement setup roughly simulates radiation emitted from a patient injected with $F^{18}$. The background radiation level (~0.08 μSv/hr) was negligible compared to the typical dose rates measured (~25 μSv/hr), and was anyhow subtracted from each measurement. The average photon energy reaching the detector was less than 511 keV, as is the case for the average photon energy emitted from a patient. For each measurement, the clock start and stop time were recorded. The source activity was measured immediately after each transmission measurement, and its value during a measurement was evaluated using the standard time dependent 1.8 hour half-life exponential decay factor.

The wall transmission coefficient for a given source-dosimeter distance d was calculated as $T=(DR)_m/(DR)_e$, where $(DR)_m$ is the measured dose rate (in Gy/sec) behind the wall, and $(DR)_e$ is the expected dose rate with the wall removed. To convert from Gy/s to μSv/h, multiply by $3.6\times10^9$, which takes into account also that Gy and Sv are equivalent for photons. $(DR)_e$ is calculated as [2],[3]:

$$(DR)e = \varepsilon * K * A / d^2 \quad (3)$$

A one Sv dose equals the absorption of one joule of photon energy per kilogram of the exposed material. Here, the source efficiency is ε, the source-detector distance is d in meters, A is the average source activity in Bq during the measurement, and K is the $F^{18}$ KERMA factor or K-Factor [3]. KERMA stands for Kinetic Energy Released per unit MAss. It can be expressed [3] as:

$$K = \frac{1}{4\pi} Y * E * \left(\frac{\mu_{en}}{\rho}\right)_{air} \quad (\text{Sv m}^2). \qquad (4)$$

The units ($m^2$) are included to account for the flux (photons per unit area). The K-factor derives from the efficiency of energy deposition in a material based on the likelihood of photon emission, the energy of those photons, the attenuation properties of the material, and its density. For $F^{18}$ decay photons in air, Y is the probability of photon emission per decay (~0.97), E is the energy of each emitted photon (0.511 MeV, converted to Joules), $\mu_{en}$ is the energy absorption coefficient (fraction of photon energy absorbed per kg of air), and ϱ is the density of air (~1.225 kg/$m^3$). Together with Eq. 3, the dose rate DR with wall removed is then determined in Sv/s. In this study, we use the air KERMA rate constant K=3.75×$10^{-17}$ Sv $m^2$/Bq s, equivalent to the value 0.134 μSv $m^2$/MBq h given by Madsen et al. [2].

The source efficiency was measured far from the wall in a horizontal position, with source-dosimeter distances d = 0.5, 1, 2, 3 m. The efficiency, taken as the average ratio between measured and expected DRs, is ε =0.35 ± 0.03. This relatively low efficiency is due to the source design, as described previously. Note that photon Compton scattering at 511 keV is forward-peaked. This angular distribution information is built into the MC simulations. It is taken into account in the experimental measurement by averaging over the different wall-detector distances.

The transmission measurements as a function of wall-detector distance, for a fixed 3.0 meter source-wall distance, are given in Table 1. The experiment measured dose rates at distances ranging from 0.05 to 3.0 m behind the wall. The source to wall distance was set at 3.0 meters. Uncertainties could be reduced in future measurements by taking more measurements, and using a dosimeter having a higher accuracy calibration.

| Wall-Detector Distance, m | Transmission Coefficient |
|---|---|
| 0.05 | 0.10 |
| 0.50 | 0.10 |
| 1.00 | 0.07 |
| 1.50 | 0.10 |
| 2.00 | 0.08 |
| 3.00 | 0.08 |

Table 1: Transmission coefficient as function of wall-detector distance, for fixed source-wall distance of 3.00 meters.

In real PET settings, personnel may be located at varying distances behind shielding walls, as close as 0.30 to 3.0 meters from the facility walls. The minimum distance 30 cm is mandated by NCRP regulation. It serves as an additional safety margin. Nevertheless, we made measurement as close as 5 cm to the shielding wall. The shielding wall must be designed to protect them from the $F^{18}$ source (the patient in the treatment room, sitting or walking). The transmission through such a protective shielding wall is generally designed to be very low (<0.5%), much lower than the ~8.86% transmission provided by 20 cm concrete. We find that the measured transmission coefficient behind the wall is roughly independent of the distance in the range studied. This shows incidentally that the contribution from air scattering is negligible.

Besides since the mfp of 511 keV photons in air is ~94 m, almost 2000 higher than the 5 cm mfp in concrete, the contribution of air scattering should be negligible through 3 or less meters of air. The independence of transmission on air scattering should therefore be the case also for walls that provide lower transmission rates. In terms of safety distance from shielding walls, our results extrapolate well to real-world radiation protection scenarios.

## 3 Theory

Stankovic et al. [4] reported a 7.0% transmission factor for 20 cm thick concrete having density 2.30 g/cm$^3$ via a MC simulation of 511 keV photons incident on an ordinary concrete cylinder of diameter 1 meter; corresponding to a B=3.2 buildup factor. Madsen et. al. [2] reported a transmission factor of 9.0% for 20 cm thick concrete having density 2.35 g/cm3 at 500 keV for an infinite broad beam geometry; corresponding to a B=4.5 buildup factor. MC uncertainties were not quoted. Although the MC and experimental geometric configurations are not identical, they are similar enough for our purposes. The mean and standard deviation of the two MC values is B = 3.9 ± 0.6, corresponding to an uncertainty of order 15%. We assume on this basis that the uncertainty for each MC simulation is of order 15%.

## 4 Results and Discussion

Table 1 showed the measured transmission coefficients through an Israeli B30 concrete [6] wall over a distance range 0.05-3.0 m. The mean and standard deviation are T = (8.8 ± 1.2)%. The value 1.2 corresponds to a 14% uncertainty. It represents mainly a statistical average of repeated measurements. The overall uncertainties then are: standard deviation: 14%; Atomex calibration: 8%; K-factor: 8%; source efficiency: 8%; positioning: 5%). No attempt is made here to separate systematic and statistical errors. We add all these errors in quadrature to give an approximate 20% total error. Therefore, the experimental wide beam transmission coefficient equals T = (8.8±1.8)%, corresponding to a B=4.0±0.8 buildup factor. This is in good agreement with Stankovic et al. T=(7.0±1.1)% and Madsen et al. T=(9.0±1.4)%; and with the mean value and standard deviation B=3.9±0.6 of these two MC simulations. Results are shown in Table 2. The values shown are appropriate only for 20 cm thickness concrete. Of course, PET facilities worldwide may use different materials; and in the case of concrete, different compositions and densities and wall thicknesses. Our objective is to compare Monte Carlo (MC) simulations with experimental values for a specific concrete composition, density, and configuration; serving as a test of the reliability of the MC calculations for our shielding construction needs. We do not aim to cover the entire range of compositions and thicknesses; instead, our results are limited to this specific setup. No attempt is made to universally validate all Monte Carlo algorithms.

All transmitted radiation doses to employees were lower than the allowed design radiation dose of 2 mSv per year [11]. The mean and standard deviation of the narrow-beam transmission coefficient measurements are T = (3.0±0.9)%. The value 0.9 corresponds to a 30% uncertainty. Following the wide-beam uncertainty analysis, adding all errors in quadrature, the experimental narrow beam transmission coefficient equals T = (3.0 ± 1.0)%, corresponding to a B=1.4±0.5 buildup factor. Considering uncertainties, this result is consistent with the T=2.0% narrow beam known value.

| Reference | Transmission Coefficient T | Buildup Factor B | Concrete Density ϱ |
|---|---|---|---|
| MC Stankovic et al. | 0.070 * | 3.2 * | 2.30 |
| MC Madsen et al. | 0.0904 * | 4.5 * | 2.35 |
| MC average | 0.08 * | 3.9±0.6 | 2.33 |
| Wide Data | 0.088±0.018 | 4.0±0.8 | 2.30 |
| Narrow Known | 0.020 | 1.0 | 2.30 |
| Narrow Data | 0.030±0.010 | 1.4±0.5 | 2.30 |

Table 2: Transmission Coefficients T for Ordinary Concrete. The Stankovic et al. and Madsen et al. results are via MC simulations, while the Data are from this experiment. The values labelled by * are taken to have uncertainties of order 15%.

Note that shielding designs based exclusively on thick concrete walls, besides being space and cost inefficient, may create construction constraints due to their weight. In addition, since ordinary concrete such as Israeli or Portland B30 has an effective atomic number around Z=10-12, 511 keV photons interact almost exclusively by one or more Compton scatterings. Transmission through such a shielding wall can be high due to a large B-factor for a very wide beam (equivalent to a moving radioactive patient) and a 20 cm thick concrete wall. When using concrete, one should consider advanced radiation shielding concrete [12],[13]. By contrast, since Z=82 for lead, scattering in a lead wall will be mainly by the photoelectric effect, which has a very high cross section. Therefore, transmission through such a wall would be very low (<0.5%) for a 3.3 cm Lead wall thickness. To achieve such low transmission with concrete would require a 40 cm thick wall. Therefore, lead is much preferred, and such lead sheets could be held in place by mounting them on a 5 cm thick gypsum wall. Thereby, the attenuation would be achieved mainly by the Lead, while the gypsum serves for physical support. Such a lead/gypsum wall is therefore much preferred to a concrete wall. It optimizes shielding design, considering space constraints and material alternatives and expense. It better protects hospital staff from a patient's radiation. Actually, considering the environmental toxicity of lead, lead-free shielding materials are preferred. Recent studies advocate the use of tungsten carbide [14],[15].

However, our primary objective was not to design an optimum shielding wall, but rather to experimentally check the results of two available MC simulations for a 20 cm thick concrete wall. These simulations and our experiment were based on wide beam assumptions. We found good agreement between MC and experimental buildup factors.

## 5 Limitations

We do not claim based on a single concrete study result that Monte Carlo (MC) simulation algorithms have been validated. Since the experimental results are specific to a particular configuration, they may not be generalized to other geometries, concrete compositions, densities, or thicknesses. In addition, our study did not include other shielding materials such as lead or tungsten. Our single study provides valuable insights and preliminary validation for the specific configuration tested. But it is insufficient to conclude that MC simulation algorithms are validated for all potential applications. Broader validation efforts are needed to establish the reliability of these algorithms comprehensively. Besides, the geometric layouts of the two MC simulations differ from that of our experiment. It would be preferred to carry out MC simulations that exactly match the experimental setup. Calibraton uncertainties could be reduced in future measurements by using a dosimeter having a higher accuracy calibration. For yet wider beam configurations, the transmission coefficient through concrete would increase even more, and such configurations merit further study. With regard to lead shielding, measurements were not carried out; and we relied only on MC simulations.

## 6 Conclusions

In this study, we have examined the protective shielding thickness requirements for concrete and lead walls in PET facilities. With a focus on the widely used $F^{18}$ radiotracer, which emits 511 keV photons, we emphasized the critical need for effective shielding to protect hospital staff during a patient's treatment and hospitalization. Our photon transmission measurements through a standard Israeli B30 concrete wall provided essential insights into the shielding performance. Our setup roughly simulates radiation emitted from a hospitalized patient injected with $F^{18}$. Our findings reveal that the narrow-beam transmission coefficient for a 20 cm thick concrete wall is T = (3.0 ± 1.0)%. By contrast, the wide-beam transmission measurement yields T = (8.8 ± 1.8)%, corresponding to B = 4.0 ± 0.8. This means that approximately four times thicker shielding would be required by a wide beam compared to a narrow beam, to achieve the same level of radiation protection. This conclusion that aligns well with the available MC simulation calculations. While our experimental results validate the MC simulations for this specific configuration, we cannot generalize this result to all shielding applications. Besides, reliance on thick concrete walls not only presents space and cost inefficiencies, but also introduces construction limitations due to increased weight. Our analysis highlights the advantages of using lead for shielding. Lead walls, with their high atomic number (Z=82), utilize the photoelectric effect, which drastically reduces photon transmission to less than 0.5% at a thickness of only 3.3 cm. Consequently, a lead shielding system, supported by a 5 cm gypsum wall, presents a superior solution that optimizes space and enhances protection against radiation. This lead-gypsum composite wall is thus a recommended option, ensuring effective staff protection and efficient resource use in hospital environments. However, considering the environmental toxicity of lead, note that recent studies advocate the use of tungsten carbide.

## 7 Declarations

***Study Limitations:*** Discussed in text

***Acknowledgments***: None

***Funding Source:*** Department of Nuclear Medicine, Sheba Medical Center, Ramat Gan, Israel

***Competing Interests:*** None

***Warning for Hazard:*** The liquid FDG F18 source is radioactive.